\documentclass[12pt]{article}
\usepackage{graphicx}
\usepackage[breaklinks]{hyperref}
\hypersetup{
    bookmarks=true,         
    unicode=false,          
    pdftoolbar=true,        
    pdfmenubar=true,        
    pdffitwindow=false,     
    pdfstartview={FitH},    
    pdftitle={My title},    
    pdfauthor={Author},     
    pdfsubject={Subject},   
    pdfcreator={Creator},   
    pdfproducer={Producer}, 
    pdfkeywords={keyword1} {key2} {key3}, 
    pdfnewwindow=true,      
    colorlinks=true,       
    linkcolor=blue,          
    citecolor=red,        
    filecolor=magenta,      
    urlcolor=cyan           
}

\newcommand\pubnumber{DPF2013-204}
\newcommand\pubdate{\today}

\def\napoli{Department of Physics\\
The Ohio State Univeristy, Columbus, Ohio, USA}

\def\Title#1{\begin{center} {\Large #1 } \end{center}}
\def\Author#1{\begin{center}{ \sc #1} \end{center}}
\def\Address#1{\begin{center}{ \it #1} \end{center}}

\newcommand\pubblock{\rightline{\begin{tabular}{l} \pubnumber\\
         \pubdate  \end{tabular}}}
\newenvironment{Abstract}{\begin{quotation}  }{\end{quotation}}
\newenvironment{Presented}{\begin{quotation} \begin{center} 
             PRESENTED AT\end{center}\bigskip 
      \begin{center}\begin{large}}{\end{large}\end{center} \end{quotation}}
\def\Acknowledgments{\bigskip  \bigskip \begin{center} \begin{large}
             \bf ACKNOWLEDGMENTS \end{large}\end{center}}

\def\beq{\begin{equation}}
\def\eeq#1{\label{#1}\end{equation}}
\def\eeqn{\end{equation}}

\def\beqa{\begin{eqnarray}}
\def\eeqa#1{\label{#1}\end{eqnarray}}
\def\eeqan{\end{eqnarray}}

\let\bar=\overbar

\def\tr{{\mbox{\rm tr}}}

\def\Dslash{\not{\hbox{\kern-4pt $D$}}}
\def\dslash{\not{\hbox{\kern-2pt $\del$}}}

\def\msb{{\bar{\ssstyle M \kern -1pt S}}}

\begin{document}
\begin{titlepage}
\pubblock

\vfill
\Title{Flavor Physics and Lattice QCD}
\vfill
\Author{ C.~M.~Bouchard}
\Address{\napoli}
\vfill
\begin{Abstract}
Our ability to resolve new physics effects is, largely, limited by the precision with which we calculate.  The calculation of observables in the Standard (or a new physics) Model requires knowledge of associated hadronic contributions.  The precision of such calculations, and therefore our ability to leverage experiment, is typically limited by hadronic uncertainties.  The only first-principles method for calculating the nonperturbative, hadronic contributions is lattice QCD.  Modern lattice calculations have controlled errors, are systematically improvable, and in some cases, are pushing the sub-percent level of precision.  I outline the role played by, highlight state of the art efforts in, and discuss possible future directions of lattice calculations in flavor physics.
\end{Abstract}
\vfill
\begin{Presented}
DPF 2013\\
The Meeting of the American Physical Society\\
Division of Particles and Fields\\
Santa Cruz, California, August 13--17, 2013\\
\end{Presented}
\vfill
\end{titlepage}
\def\thefootnote{\fnsymbol{footnote}}
\setcounter{footnote}{0}

\section{Introduction}
I begin with a discussion of the role lattice QCD plays in flavor physics, including an outline of the steps in a typical lattice simulation and the errors involved.  
I then discuss so called ``gold plated" quantities by way of three examples of state of the art calculations.
Next, I discuss lattice efforts pushing the boundaries of what can be done, and end with some possible future directions in lattice flavor physics.
I focus on quark (as opposed to lepton) flavor physics, as this is where most lattice flavor physics work is done.
Muon $g-2$, addressed in Sec.~\ref{sec-envelope}, is a notable exception.

\section{The Role of Lattice QCD}
There are two disparate energy scales in quark flavor physics processes -- the low energy scale of hadronization, $\Lambda_{\rm QCD}= {\it few} \times100$ MeV, and the high energy scale at which the electroweak interactions occur, $\Lambda_{\rm EW} \sim 100$ GeV.
As we pit experiment vs. theory, we are aided on the theory side by the disparity of these scales.  Under the Operator Product Expansion~\cite{OPE1, OPE2}, 
\begin{equation}
{\rm observable} = \sum_i {\color{black}C_i(\mu)}\ {\color{black}{\rm ME}_i(\mu)}\ +\ \mathcal{O}\left( \frac{\Lambda_{\rm QCD}}{\Lambda_{\rm EW}} \right)^2,
\end{equation}
physics associated with these scales factorizes.  The high energy, short distance physics of the electroweak interaction is contained in the Wilson coefficients {\color{black}$C_i(\mu)$}, which are generally perturbatively calculable.  The low energy, long distance physics of hadronization is contained in the hadronic matrix elements {\color{black}${\rm ME}_i(\mu)$}.  
The hadronic matrix elements are nonperturbative, and the only first-principles method for computing them is lattice QCD.
The difference in scales results in a clean separation, with corrections to the leading order factorization of about 1 part in $10^{5}$.

\subsection{Lattice Basics}

A typical lattice calculation extracts physics quantities from correlation functions.
\begin{equation}
\langle O \rangle = \frac{\int [dG]\ O[G, (\slash{\!\!\!\!D}+m)^{-1}]\ e^{-S[G] + \ln \det(\slash{\!\!\!\!D}+m)}}{ \int [dG]\ e^{-S[G] + \ln \det(\slash{\!\!\!\!D}+m)} }
\label{eq-vev}
\end{equation}
Eq.~(\ref{eq-vev}) writes a correlation function, in euclidean path integral representation, as the vacuum expectation value of composite operator $O$.  QCD gauge fields are represented by $G$ and quark masses by $m$.
A euclidean metric exponentially suppresses quantum contributions.  This suppression in Minkowski metric occurs via cancellations from the rapidly varying phase of $e^{iS}$ and is computationally difficult to implement.
Integrating quark fields by hand avoids having to work with Grassmann variables on the computer.
Operator $O$ creates particles and inserts interactions we wish to study.  Consider the semileptonic decay $B\to \pi \bar{l}\nu$.  We create a $B$ meson at time 0, insert a $b\to u$ flavor-changing current at $t>0$, then annihilate a pion at $T>t$,
\begin{eqnarray}
O &=& (\bar{u} \gamma_5 d)_T\ (\bar{b} \gamma_\mu u)_t\ (\bar{d} \gamma_5 b)_0 \nonumber \\
&=& \tr\{ (\slash{\!\!\!\!D}+m_u)^{-1}_{t,T}\ \gamma_5\ (\slash{\!\!\!\!D}+m_d)^{-1}_{T,0}\ \gamma_5\ (\slash{\!\!\!\!D}+m_b)^{-1}_{0,t}\ \gamma_\mu \}. \label{eq-wick}
\end{eqnarray}
Quark fields are Wick contracted and $O$ is written in terms of quark propagators and gamma matrices.  Eqs.~(\ref{eq-vev}) and~(\ref{eq-wick}) represent the ``jumping off point" for a lattice simulation.  We discretize the action, replacing $\int d^4x \to \sum_{x,y,z,t}$, writing $\slash{\!\!\!\!D}$ as a finite difference, and \emph{much} more (which I gloss over here).  After discretization, the path integral can be evaluated with Monte Carlo methods and a collection of gauge field configurations $\{G\} = \{G_n, n=1\dots N \}$ generated with probability distribution $e^{-S[G] + \ln \det(\slash{\!\!\!\!D}+m)}$.  The path integral in eq.~(\ref{eq-vev}) is then simply the average of $O$ on $\{G\}$,
\begin{equation}
\langle O \rangle^{\rm latt}(a) = \frac{1}{N} \sum_{n=1}^{N} O[G_n, (\slash{\!\!\!\!D}+m)^{-1}_n].
\end{equation}
Quantities obtained from $\langle O \rangle^{\rm latt}(a)$, unless they are renormalization group invariant (RGI), are specific to a lattice regulator and scale $\sim\!a^{-1}$.
We connect quantities calculated on the lattice to a continuum scheme and scale $\mu$ with a matching factor $Z(\mu,a)$, typically determined from lattice perturbation theory~\cite{LPT},
\begin{equation}
\langle O \rangle(\mu) = Z(\mu,a)\ \frac{1}{N} \sum_{n=1}^{N} O[G_n, (\slash{\!\!\!\!D}+m)^{-1}_n].
\label{eq-err}
\end{equation}
The simulation is repeated at multiple lattice spacings and quark masses to allow extrapolation to the continuum and to physical quark mass.  I gloss over the subsequent steps of fitting simulation data to extract physics quantities and extrapolating these quantities to the continuum and to physical quark masses, and refer the interested reader to ref.~\cite{DeTar}.

\subsection{Lattice Errors}

Eq.~(\ref{eq-err}) highlights many of the typical errors in a lattice calculation:
\begin{itemize}

\item {\bf matching}:  The error associated with the determination of $Z(\mu,a)$, often done to one loop in lattice perturbation theory, can be improved by working to higher order or using nonperturbative methods (e.g. refs.~\cite{nonpert1, nonpert2}), or eliminated altogether if an appropriate RGI matrix element can be identified.

\item {\bf statistical}:  Uncertainty associated with the Monte Carlo evaluation of the path integral goes like $N^{-1/2}$, and can be improved with brute force by increasing $N$.

\item {\bf discretization}:  Replacing continuum quantities with their discrete counterparts introduces corrections of $\mathcal{O}(a^n)$, where $n$ depends on the discretization.  
These errors are reduced with smaller lattice spacing and lattices as fine as \mbox{0.03 fm} are currently being generated~\cite{charm}.
Explicitly accounting for discretization effects can also reduce this error, and increased statistics generally make it easier to identify discretization effects in the data.

\item {\bf input parameters}:  Errors due to the uncertainty of physical parameters, like masses, that get ``plugged in" to a calculation are typically $<1\%$.

\item {\bf scale setting}:  Similar to input parameters, scale setting refers to determining the lattice spacing in fm.  It is accomplished by matching a lattice quantity to its known value.  This quantity could be a standard observable like a hadron mass, or a more theoretical object, e.g. ref.~\cite{r1}.  This is an active area of research and new methods are under development~\cite{flow}.  Whatever quantity is used, its lattice determination generally improves with increased statistics.

\item {\bf correlation function fits}:  Extracting physics quantities from lattice simulation data is a significant analysis step.  
Increased statistics helps, but systematic errors must also be accounted for.  For this, Bayesian fitting~\cite{fits} is useful.

\item {\bf chiral extrapolation}:  Guided by chiral perturbation theory, correlation function fit results are extrapolated to physical light quark mass and infinite volume.  
These extrapolations generally improve with statistics and the extrapolation in mass can be eliminated altogether by simulating at the physical light quark mass.

\item {\bf EM, isospin breaking, and charm sea effects}:  Simulations often omit QED, assume degenerate up and down quarks, and neglect effects from charm and heavier sea quarks.  These effects, typically $<1\%$, are now being included in high precision simulations~\cite{charm, EM1, EM2}.

\end{itemize}

\section{``Gold Plated" Processes}
Processes relatively easy to calculate on the lattice, usually characterized by no more than one initial ground state hadron connected by a local interaction to no more than one final ground state hadron, are called ``gold plated".
The study of gold plated processes that occur at tree-level in the SM, ie. leptonic and semileptonic decays, allow precision determinations of CKM matrix elements and test the SM accommodation of quark flavor-changing interactions.
%
Studying gold plated rare processes, ie. rare decays or short distance contributions to meson mixing, provides useful constraints on new physics.
I discuss three examples of recent lattice calculations that illustrate state of the art lattice efforts involving (i) pions and kaons in SM tree-level processes, (ii) $B$ mesons in SM tree-level processes, and (iii) rare processes.
Time limitations prevent me from also discussing the determination of SM parameters (quark masses and $\alpha_s$).  I refer the reader to refs.~\cite{FLAG, McNeile}, which summarize recent lattice results.

\begin{figure}[t]
\hspace{-0.3in}
\includegraphics[height=2.6in]{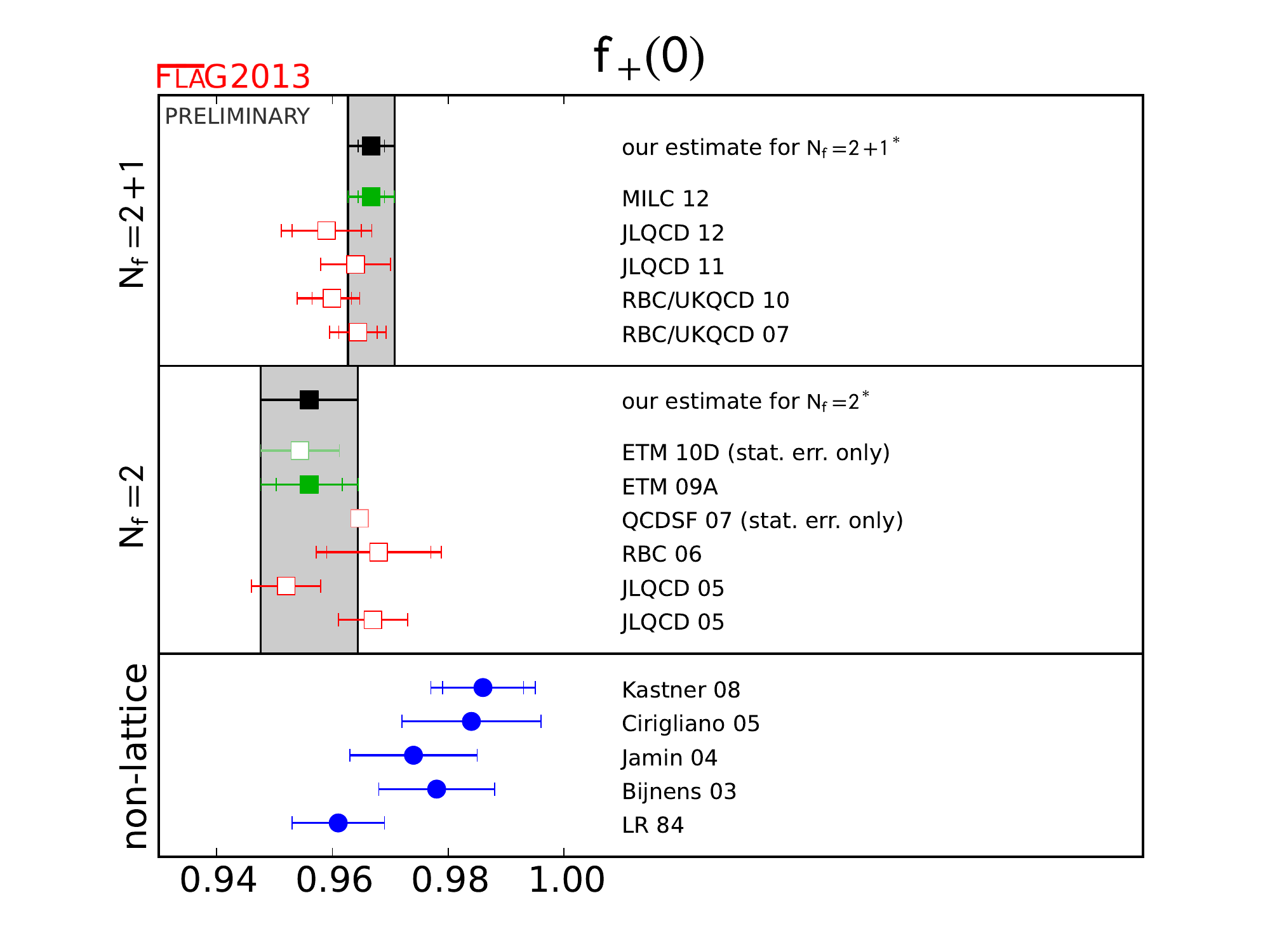}
\hspace{-0.3in}
\includegraphics[height=2.6in]{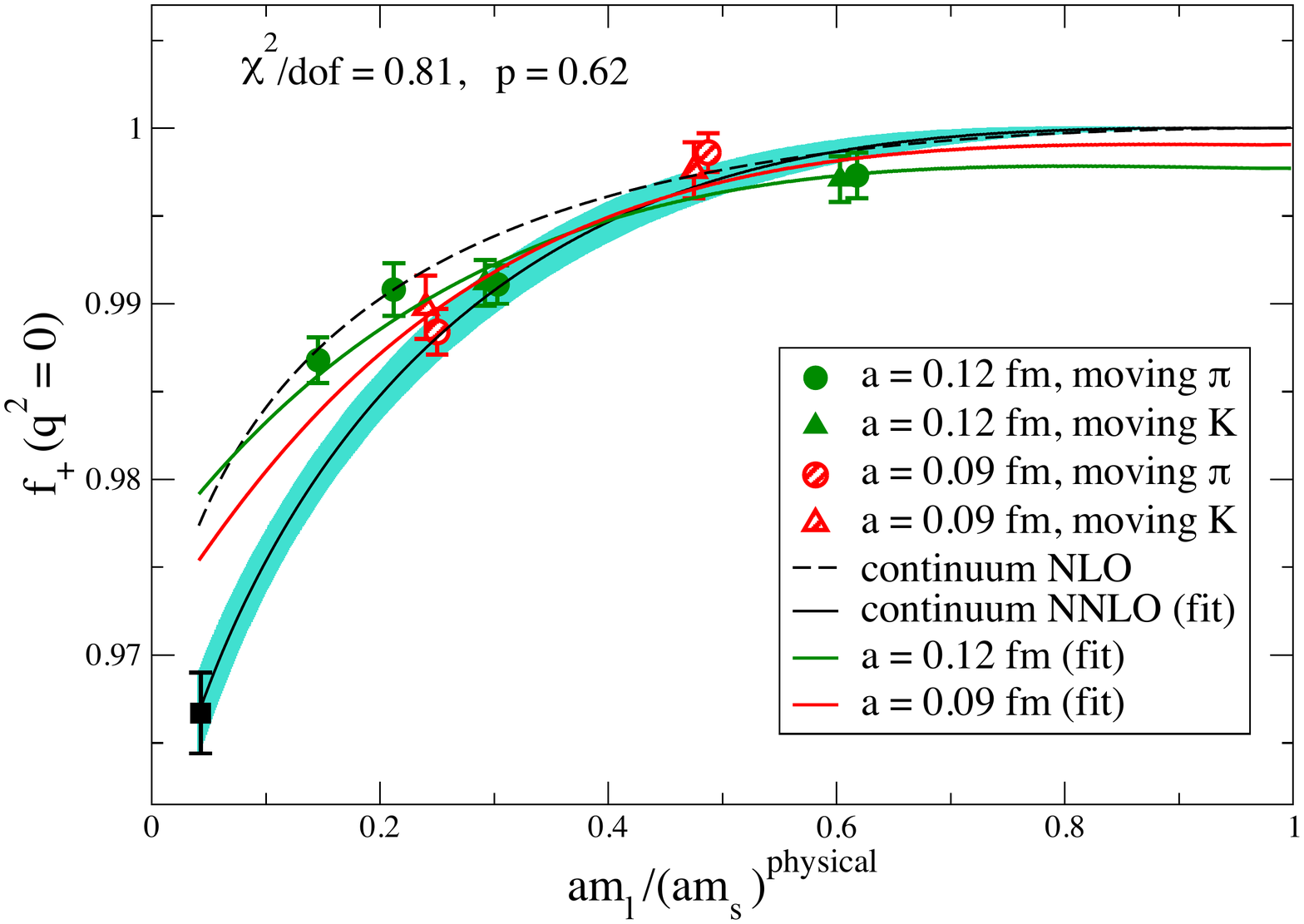}
\caption{(left) FLAG~\cite{FLAG} summary of results.  The result of ref.~\cite{Gamiz1} is labeled ``MILC 12".  (right) The chiral and continuum extrapolation of ref.~\cite{Gamiz1}.}
\label{fig:fplus}
\end{figure}
\subsection{Pions and Kaons with Sub-Percent Precision}
The semileptonic decay rate $\Gamma(K\to \pi l \bar\nu)$ is related to the hadronic vector form factor $f_+(0)$, and CKM matrix element $|V_{us}|$ by~\cite{Cirig}
\begin{equation}
\frac{ \Gamma(K\to\pi l \bar\nu) }{ I^{(0)}_{Kl}S_{\rm EW} (1+\delta_{\rm EM}^{Kl} + \delta_{\rm SU(2)}^{K\pi}) } = |V_{us}|^2\  f_+^2(0),
\label{eq-Kpi}
\end{equation}
where the lhs of eq.~(\ref{eq-Kpi}), the measured decay rate with known corrections, is known to $0.2\%$.  Our ability to leverage this precision in a determination of $|V_{us}|$ is limited by the precision of $f_+(0)$.
Ref.~\cite{Gamiz1} uses the kinematic constraint $f_+(0)=f_0(0)$ to recast $f_+(0)$ in terms of the absolutely normalized scalar matrix element
\begin{equation}
f_0(q^2) = \frac{ m_s - m_l }{ M_K^2 - M_\pi^2 } \langle \pi | S^{\rm latt} | K \rangle_{q^2}.
\end{equation}
This eliminates the matching error and results in a determination of $f_+^2(0)$ with $0.8\%$ error.
A comparison with other results is shown in Fig.~\ref{fig:fplus}.
Despite this impressive precision, there is still room for improvement.  
A follow on effort~\cite{Gamiz2} is addressing the leading source of error, the chiral extrapolation (also shown in Fig.~\ref{fig:fplus}), by simulating at the physical light quark mass.  An error of about $0.6\%$ is anticipated.

\begin{figure}[t]
\hspace{-0.3in}
\includegraphics[height=2.6in]{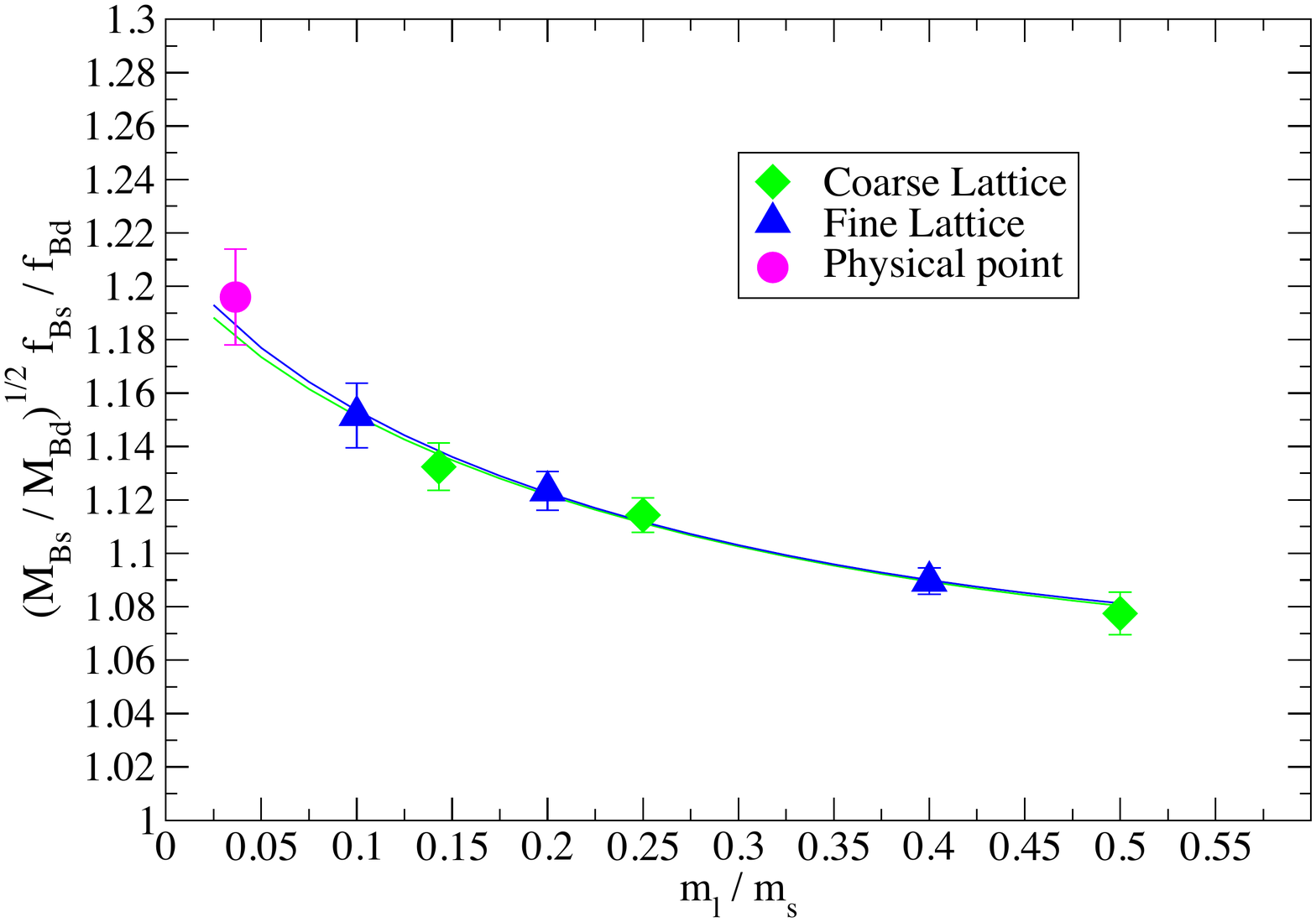}
\hspace{-0.3in}
\includegraphics[height=2.6in]{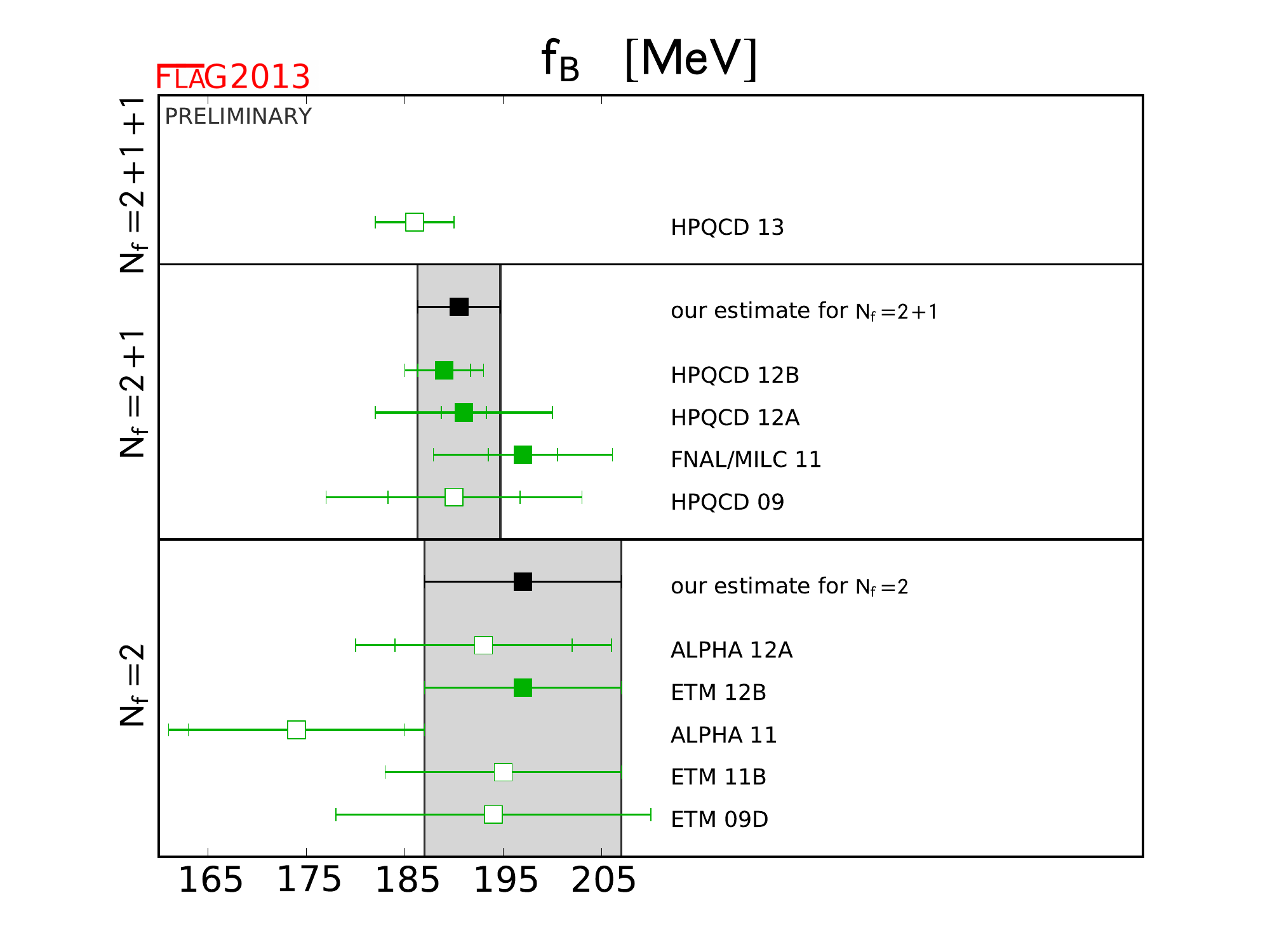}
\caption{(left) Chiral and continuum extrapolation of $f_{B_s}/f_B$~\cite{ratio}.  (right) The FLAG~\cite{FLAG} summary of $f_B$ results.  The result of ref.~\cite{ratio} is labeled ``HPQCD 12B".}
\label{fig:fB}
\end{figure}
\subsection{$B$ Mesons with 2\% Precision}
The precision achieved in $K\to\pi l\bar\nu$ comes, in large part, from the absence of a matching error due to clever choice of matrix element.  If an effective theory treatment had been used for, say, the strange quark, a matching factor would be needed to match the effective and physical theories.  This is the situation in $B$ physics, where the combination of a heavy $b$ and light ($u$ or $d$) valence quark requires both large volume \emph{and} fine lattices --- this is prohibitively expensive\footnote{A light quark has a long Compton wavelength, requires large volumes, and is expensive to simulate.  Conversely, a heavy quark requires a finer lattice to resolve its short Compton wavelength.}.  A common solution is the use of an effective theory (e.g. Nonrelativistic QCD) for the $b$ quark with a resultant {\it few} \% matching error.
For the decay constant $f_B$, which characterizes the hadronic contribution to \mbox{$B\to \bar l\nu$}, a clever way around this dilemma was developed by HPQCD.  In~\cite{ratio} they calculated the ratio $f_B/f_{B_s}$ using Nonrelativistic QCD for the $b$ quark.  In the ratio, the {\it few} \% matching error all but cancels, giving a determination with 1.5\% error.  The problems of simulating a heavy $b$ and a light ($u$ or $d$) quark are not present when the light quark is replaced by the heavier $s$ quark.  In~\cite{fBs} they calculated $f_{B_s}$, sans matching error, without resorting to effective theory for the $b$ quark.  Combining $f_B/f_{B_s}$ and $f_{B_s}$ gives a 2.1\% determination of $f_B$~\cite{ratio}.
In~\cite{newfB}, HPQCD simulates at the physical light quark mass, uses an improved effective theory for the $b$ quark, and estimates a small matching error.  Their result, labeled ``HPQCD 13" in Fig.~\ref{fig:fB}, has a similarly small error.

\begin{figure}[t]
\hspace{-0.5in}
\includegraphics[angle=-90, totalheight=2.65in]{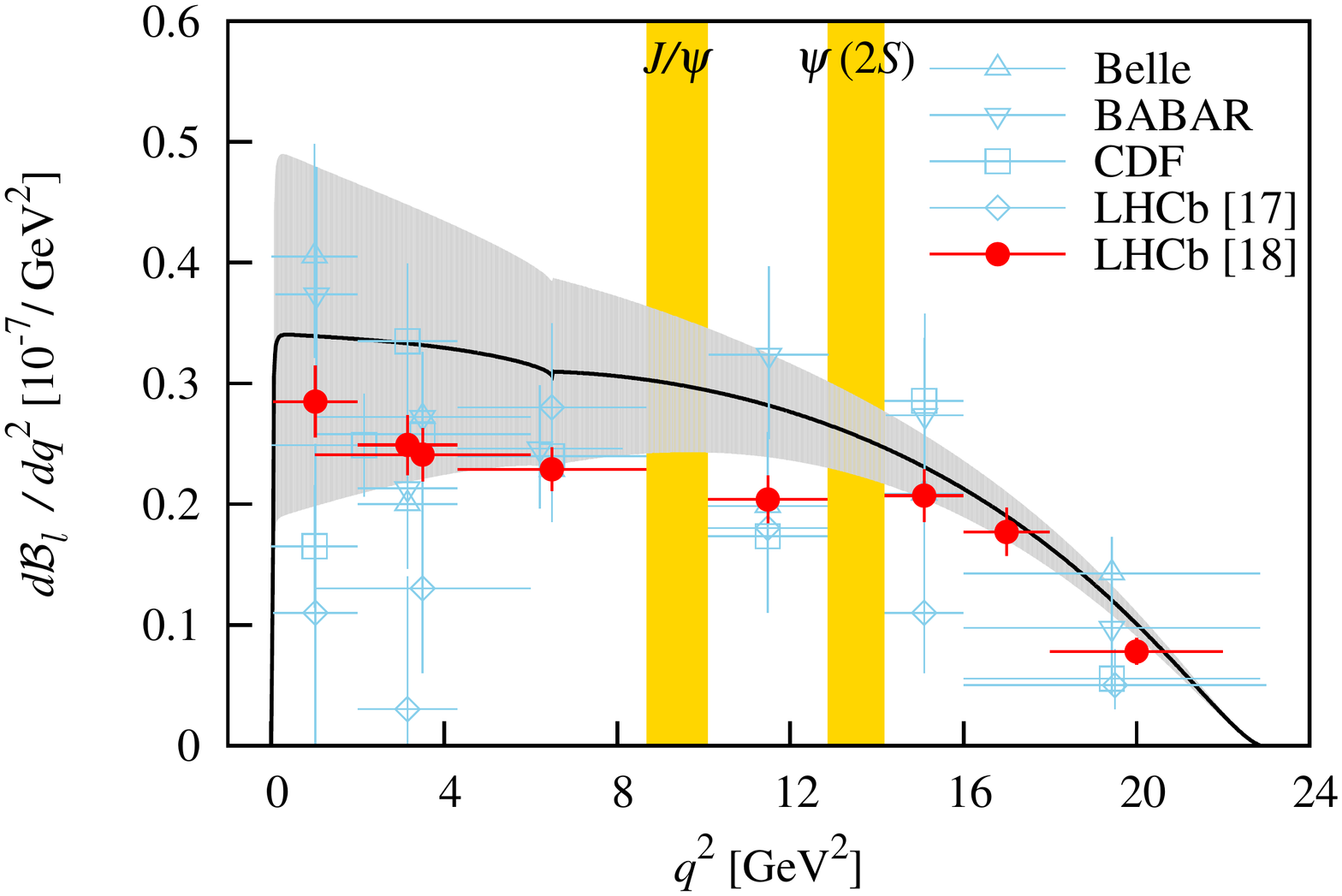}
\hspace{-0.3in}
\includegraphics[angle=-90, totalheight=2.65in]{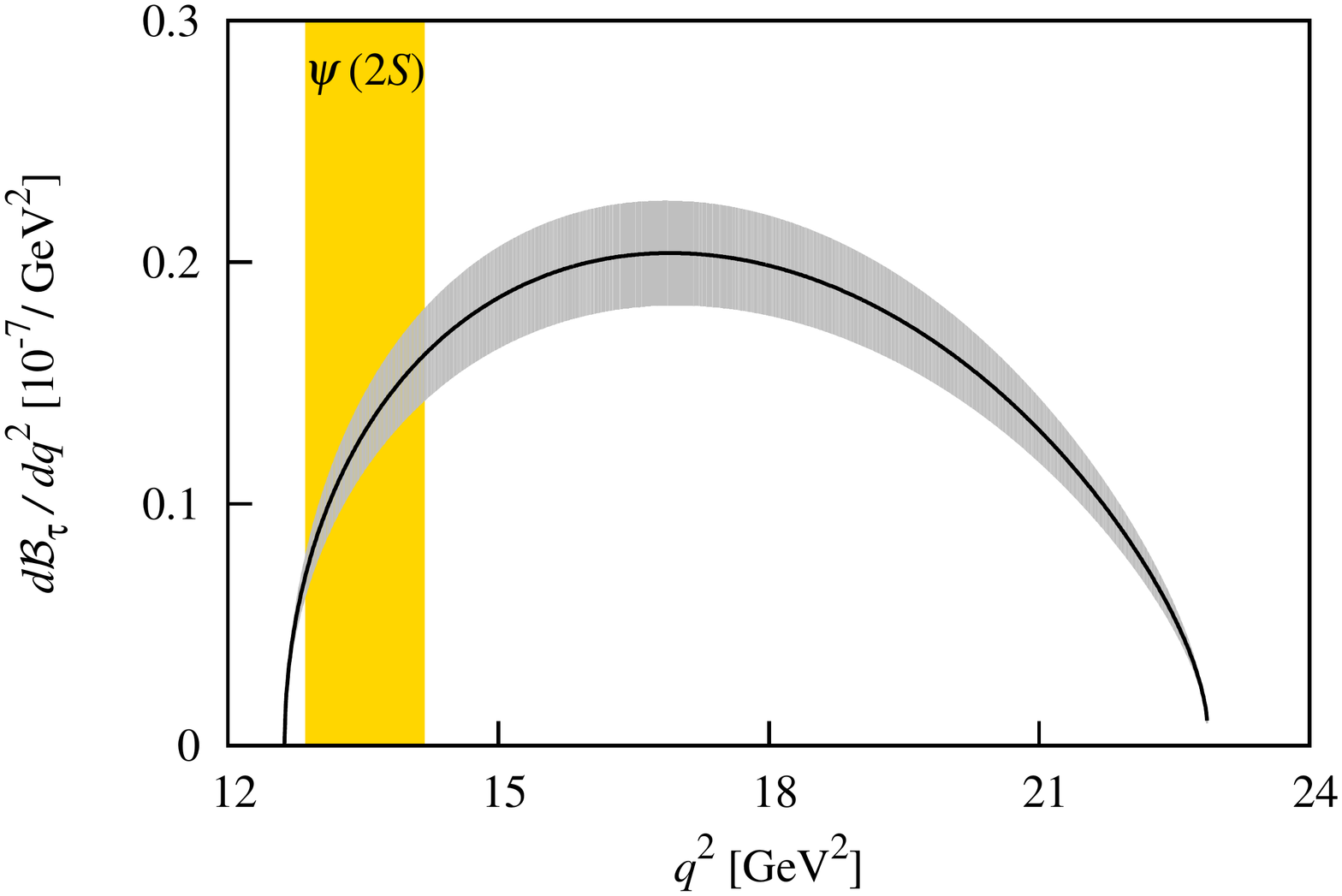}
\vspace{-0.2in}
\caption{$B\to K\ell\bar\ell$ differential branching fractions for (left) light dilepton and (right) ditau final states.}
\label{fig:BtoK}
\end{figure}
\subsection{Rare Processes}
The FCNC's that mediate these processes require loop diagrams in the SM, opening the door for possible detectable new physics effects.  The decay $B\to K \ell\bar\ell$ proceeds via a $b\to s$ transition and is used, often in combination with other rare decays, to constrain new physics models.  In~\cite{PRD} HPQCD calculates the hadronic contributions relevant to the SM and generic new physics models and in~\cite{PRL} uses these results to make SM predictions and compare with experiment.  Fig.~\ref{fig:BtoK} shows the SM predictions, and experiment where available, versus the squared four-momentum of the leptons.  The SM branching fraction for light dileptons is consistent with experiment.  The silver lining: this leaves little room for new physics effects and generates stronger constraints on new physics models.  The possibility of new physics related to the tau motivates measurements of $B\to K\tau\bar\tau$.


\section{Pushing the Envelope}
\label{sec-envelope}
Important advances in lattice QCD are expanding its influence to new regions of flavor physics.  I briefly discuss a few of these efforts.  Additional examples can be found in ref.~\cite{ProjX}.

Despite notoriously difficult long distance effects from intermediate states, impressive progress is being made for the $K_L-K_S$ mass difference $\Delta M_K$ in kaon mixing~\cite{longd}.  Accurate first principles results for $\Delta M_K$, and perhaps the indirect CP violation parameter $\epsilon_K$, where additional difficulties arise, are expected in the coming years.

Processes with multiple final state hadrons are complicated by finite volume effects, final state interactions, and other issues.  A systematic treatment of these complications is underway for $K\to\pi\pi$ with exciting results~\cite{Kpipi}.  Heavier initial states add further complications as additional multiparticle channels open.  However, promising first steps have been made for $D\to\pi\pi$ and $D\to K\bar K$~\cite{Dpipi}.

Leveraging anticipated improvement in the experimental precision of the muon anomalous magnetic moment will require an improved understanding of hadronic contributions.  There have been impressive advances in lattice calculations of the hadronic vacuum polarization~\cite{HVP1, HVP2, HVP3, HVP4, HVP5, HVP6} and pioneering work on the much harder hadronic light by light contribution~\cite{HLbL1, HLbL3, HLbL4}.

Calculations of quasielastic neutrino-nucleon scattering, needed to interpret neutrino oscillation experiments, parameterize hadronic nuclear effects via the nucleon axial form factor $F_A(Q^2)$.  Lattice calculations~\cite{FA1, FA2} and model-independent parameterizations of the $Q^2$ dependence~\cite{FA3} are essential to obtain first principles results.

Neutron oscillation~\cite{neut} and proton decay~\cite{pdecay} matrix elements have been calculated on the lattice.  When combined with (non)observation, these matrix elements can provide stringent tests of new physics models.

\section{The Future}
I end with speculation about possible future lattice QCD efforts in flavor physics.  

Gold plated quantities will continue to be pushed toward 1\% precision. 
Reaching this level of precision for all gold plated observables will require computational and algorithmic advances that permit simulations with physical light and heavy quark masses, and using lattices fine enough and volumes large enough that associated systematic errors are negligible.  Nonperturbative, or higher order perturbative, calculations of matching factors will be needed along with continued improvements in scale setting and an accounting of QED effects.

First principles calculations of form factors for decays to final state hadrons unstable in QCD, e.g. $B\to K^*\ell\bar\ell$ and $B\to\rho \ell\bar\nu$, are needed in the search for new physics.
To fully leverage experimental results, the calculations should account for subsequent decays without introducing approximations accompanied by uncontrolled errors~\cite{NWA}.

Calculations of long distance contributions to $D$ mixing and the rare decay \mbox{$K\to\pi\ell\bar\ell$} are motivated by theory and experiment.  Given their phenomenological importance, attempts to understand these long distance effects are likely, perhaps building on the methods developed in~\cite{longd} for kaon mixing.

If extensions of~\cite{Dpipi} are successful in dealing with channels including more than two pions in the nonleptonic weak decays $D\to\pi\pi$ and $D\to K\bar K$, it is tempting to extrapolate to the $B$ meson and speculate about, e.g., lattice studies of CP violation in $B\to \pi\pi$ and $B_s\to KK$.

\Acknowledgments
I thank the conveners for the invitation to speak, and the organizers for an enjoyable conference.


\begin{thebibliography}{99}


\bibitem{OPE1}
K.G.~Wilson,
Phys. Rev. {\bf 179}, 1499 (1969)
[\href{http://prola.aps.org/abstract/PR/v179/i5/p1499_1}{doi:10.1103/PhysRev.179.1499}].

\bibitem{OPE2}
K.G.~Wilson and W.~Zimmerman,
Commun. math. Phys. {\bf 24}, 87 (1972)
[\href{http://link.springer.com/article/10.1007\%2FBF01878448}{doi:10.1007/BF01878448}].

\bibitem{LPT}
A well organized collection of lattice perturbation theory references can be found at \mbox{\href{http://latticeperturbationtheory.org/}{latticeperturbationtheory.org}}.

\bibitem{DeTar}
T.~DeGrand and C.~DeTar,
{\it Lattice Methods for Quantum Chromodynamics} 
(World Scientific, New Jersey, 2006).

\bibitem{nonpert1}
C.~Sturm et al., Phys. Rev. {\bf D80}, 014501 (2009) 
[\href{http://arxiv.org/abs/0901.2599}{arXiv:0901.2599}].

\bibitem{nonpert2}
Y.~Aoki et al., Phys. Rev. {\bf D84}, 014503 (2011) 
[\href{http://arxiv.org/abs/1012.4178}{arXiv:1012.4178}].

\bibitem{charm}
A.~Bazavov et al. (MILC),
Phys. Rev. {\bf D87}, 054505 (2013)
[\href{http://arxiv.org/abs/1212.4768}{arXiv:1212.4768}].


\bibitem{r1}
C.~Aubin et al.,
Phys. Rev. {\bf D70}, 094505 (2004)
[\href{http://arxiv.org/abs/hep-lat/0402030}{hep-lat/0402030}].

\bibitem{flow}
M.~L\"uscher, 
\href{http://arxiv.org/abs/arXiv:1308.5598}{arXiv:1308.5598}.

\bibitem{fits}
G.P.~Lepage et al.,
Nucl. Phys. Proc. Suppl. {\bf106}, 12 (2002)
[\href{http://arxiv.org/abs/hep-lat/0110175}{hep-lat/0110175}].

\bibitem{EM1}
T.~Ishikawa et al., 
Phys. Rev. Lett. {\bf 109}, 072002 (2012)
[\href{http://arxiv.org/abs/1202.6018}{arXiv:1202.6018}].

\bibitem{EM2}
S.~Aoki et al. (PACS-CS), 
Phys. Rev. {\bf D86}, 034507 (2012)
[\href{http://arxiv.org/abs/1205.2961}{arXiv:1205.2961}].


\bibitem{FLAG}
The Flavor Lattice Averaging Group provides a fairly comprehensive list of averaged lattice results at \mbox{\href{http://itpwiki.unibe.ch/flag}{http://itpwiki.unibe.ch/flag}}.

\bibitem{McNeile}
C.~McNeile,
Mod. Phys. Lett. {\bf A28}, 1360012 (2013) 
[\href{http://arxiv.org/abs/1306.3326}{arXiv:1306.3326}].

      
\bibitem{Cirig}
V.~Cirigliano et al., 
Rev. Mod. Phys.{\bf  84}, 399 (2012)
[\href{http://arxiv.org/abs/1107.6001}{arXiv:1107.6001}].

\bibitem{Gamiz1}
E.~G\'amiz et al.
(Fermilab Lattice and MILC),
Phys. Rev. {\bf D87}, 073012 (2012)
[\href{http://arxiv.org/abs/1212.4993}{arXiv:1212.4993}].

\bibitem{Gamiz2}
E.~G\'amiz et al.
(Fermilab Lattice and MILC),
talk at Lattice 2013.

\bibitem{ratio}
H.~Na et al.
(HPQCD),
Phys. Rev. {\bf D86}, 034506 (2012)
[\href{http://arxiv.org/abs/1202.4914}{arXiv:1202.4914}].

\bibitem{fBs}
C. McNeile et al.
(HPQCD),
Phys. Rev. {\bf D85}, 031503 (2012)
[\href{http://arxiv.org/abs/1110.4510}{arXiv:1110.4510}].

\bibitem{newfB}
R.J.~Dowdall et al.
(HPQCD),
Phys. Rev. Lett. {\bf 110}, 222003 (2013)
[\href{http://arxiv.org/abs/1302.2644}{arXiv:1302.2664}].

\bibitem{PRD}
C.M.~Bouchard et al.
(HPQCD),
to appear in Phys. Rev. D
[\href{http://arxiv.org/abs/1306.2384}{arXiv:1306.2384}].

\bibitem{PRL}
C.M.~Bouchard et al.
(HPQCD),
to appear in Phys. Rev. Lett.
[\href{http://arxiv.org/abs/1306.0434}{arXiv:1306.0434}].

\bibitem{ProjX}
A.S.~Kronfeld and R.~Tschirhart (editors), et al.,
\href{http://arxiv.org/abs/1306.5009}{arXiv:1306.5009}.

\bibitem{longd}
N.H.~Christ et al.,
\href{http://arxiv.org/abs/1212.5931}{arXiv:1212.5931}.

\bibitem{Kpipi}
C.~Kelly's talk at this conference.

\bibitem{Dpipi}
M.T.~Hansen and S.R.~Sharpe,
Phys. Rev. {\bf D86}, 016007 (2012)
[\href{http://arxiv.org/abs/1204.0826}{arXiv:1204.0826}].

\bibitem{Liu}
Z.~Liu et al.,
\href{http://arxiv.org/abs/1101.2726}{arXiv:1101.2726}.

\bibitem{neut}
M.I.~Buchoff et al.,
PoS {\bf LATTICE2012} 128
[\href{http://arxiv.org/abs/1207.3832}{arXiv:1207.3832}].

\bibitem{pdecay}
Y.~Aoki et al.,
\href{http://arxiv.org/abs/1304.7424}{arXiv:1304.7424}.

\bibitem{HVP1}
T.~Blum,
Phys. Rev. Lett. {\bf 91}, 052001 (2003)
[\href{http://arxiv.org/abs/hep-lat/0212018}{hep-lat/0212018}].

\bibitem{HVP2}
M.~G\"ockeler et al.
(QCDSF),
Nucl. Phys. {\bf B688}, 135 (2004)
[\href{http://arxiv.org/abs/hep-lat/0312032}{hep-lat/0312032}].

\bibitem{HVP3}
C.~Aubin and T.~Blum,
Phys. Rev. {\bf D75}, 114502 (2007)
[\href{http://arxiv.org/abs/hep-lat/0608011}{hep-lat/0608011}].

\bibitem{HVP4}
X.~Feng et al.
(ETMC),
Phys. Rev. Lett. {\bf 107}, 081802 (2011)
[\href{http://xxx.lanl.gov/abs/1103.4818}{arXiv:1103.4818}].

\bibitem{HVP5}
P.~Boyle et al.,
Phys. Rev. {\bf D85}, 074504 (2012)
[\href{http://arxiv.org/abs/1107.1497}{arXiv:1107.1497}].

\bibitem{HVP6}
M.~Della~Morte et al.,
JHEP {\bf 1203}, 055 (2012)
[\href{http://arxiv.org/abs/1112.2894}{arXiv:1112.2894}].

\bibitem{HLbL1}
M.~Hayakawa et al., 
PoS {\bf LATTICE 2005} 353
[\href{http://arxiv.org/abs/hep-lat/0509016}{hep-lat/0509016}].


\bibitem{HLbL3}
T.~Blum et al.,
PoS {\bf LATTICE 2012} 022
[\href{http://arxiv.org/abs/1301.2607}{arXiv:1301.2607}].

\bibitem{HLbL4}
X.~Feng et al.
(JLQCD),
Phys. Rev. Lett. {\bf 109}, 182001 (2012)
[\href{http://arxiv.org/abs/1206.1375}{arXiv:1206.1375}].

\bibitem{FA1}
S.~Capitani et al.,
Phys. Rev. {\bf D86}, 074502 (2012)
[\href{http://arxiv.org/abs/1205.0180}{arXiv:1205.0180}].

\bibitem{FA2}
R.~Horsley et al.
(QCDSF),
\href{http://arxiv.org/abs/1302.2233}{arXiv:1302.2233}.

\bibitem{FA3}
B.~Bhattacharya et al.,
Phys. Rev. {\bf D84}, 073006 (2011)
[\href{http://arxiv.org/abs/1108.0423}{arXiv:1108.0423}].


\bibitem{NWA}
D.~Berdine et al.,
Phys. Rev. Lett. {\bf 99}, 111601 (2007)
[\href{http://arxiv.org/abs/hep-ph/0703058}{hep-ph/0703058}].


\end{thebibliography}
\end{document}